\documentclass[twocolumn,showpacs,floats,floatfix,superscriptaddress,aps,pra]{revtex4-1}
\usepackage{amsfonts}
\usepackage{amssymb}
\usepackage{amsmath}
\usepackage{graphicx}
\usepackage{bm}
\usepackage{color}
\usepackage{epsfig}
\usepackage{calc}
\usepackage{ifthen}
\usepackage{subfigure}
\usepackage{graphicx}
\usepackage{epstopdf}
\usepackage{epsfig}

\begin{document}

\title{Analogue of collectively induced transparency in metamaterial}
\date{\today }

\begin{abstract}
Most recently, a brand new optical phenomenon, collectively induced transparency (CIT) has already been proposed in the cavity quantum electrodynamics system, which comes from the coupling between the cavity and ions and the quantum interference of collective ions. In this paper, we propose the CIT in terahertz (THz) metamaterial device by employing the coupling between bright mode and interference of dark modes for the first time. We give the theoretical analysis, analytical calculations and simulations to present the transmission spectrum of CIT metamaterials. Furthermore, we can observe the tendency of CIT's transmission spectrum by experiments which well verify our idea. Ideal CIT metamaterial device can produce a very high Q peak in the middle of transmission spectrum of Electromagnetically induced transparency (EIT), which can be useful for highly sensitive metamaterial sensors, optical switches and photo-memory.

\end{abstract}

\author{Wei Huang}
\affiliation{Guangxi Key Laboratory of Optoelectronic Information Processing, School of Optoelectronic Engineering, Guilin University of Electronic Technology, Guilin 541004, China}

\author{Shi-Ting Cao}
\affiliation{Department of Electrical and Electronic Engineering, Southern University of Science and Technology, Shenzhen, 518055 China}
\affiliation{Guangxi Key Laboratory of Optoelectronic Information Processing, School of Optoelectronic Engineering, Guilin University of Electronic Technology, Guilin 541004, China}

\author{Shi-Jun Liang}
\affiliation{Institute of Brain-Inspired Intelligence, National Laboratory of Solid State Microstructures, School of Physics, Collaborative Innovation Center of Advanced Microstructures, Nanjing University, Nanjing, China}

\author{Shan Yin}
\email{syin@guet.edu.cn}
\affiliation{Guangxi Key Laboratory of Optoelectronic Information Processing, School of Optoelectronic Engineering, Guilin University of Electronic Technology, Guilin 541004, China}

\author{Wentao Zhang}
\email{zhangwentao@guet.edu.cn}
\affiliation{Guangxi Key Laboratory of Optoelectronic Information Processing, School of Optoelectronic Engineering, Guilin University of Electronic Technology, Guilin 541004, China}

\maketitle



\section{Introduction}
Most recently, collectively induced transparency (CIT) has already been proposed in the cavity quantum electrodynamics system \cite{Lei2023}, which produces the very shape peak of the reflection spectrum and it can be very useful for quantum sensor devices, quantum memory devices and quantum information processing. CIT comes from the coupling between the cavity and a large number of ions and quantum interference of collective ions. Due to the different quantum interference of collective ions by varying excitation of external laser and then coupling between the cavity, the reflection spectrum can vary from dipole induced reflectivity (DIR) \cite{Waks2006} to CIT. Thus,  the fundamental idea of CIT is different interaction between the cavity and collective ions by varying quantum interference of collective ions.

\begin{figure}[htbp]
	\centering
	\includegraphics[width=0.45\textwidth]{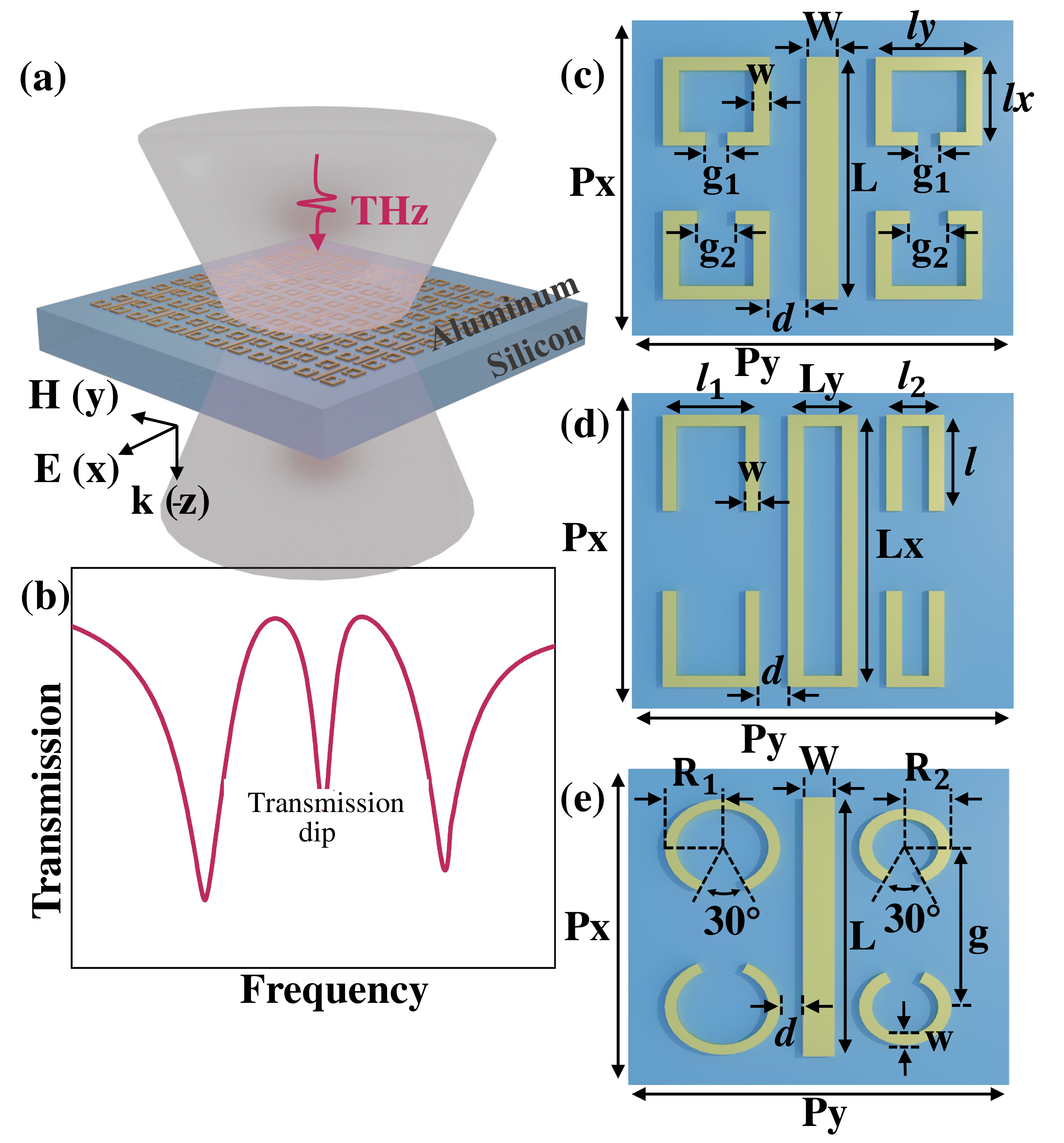}
	\caption{(a) The schematic figure of material collectively induced transparency (CIT) with transmission spectrum. (b) The schematic figure of CIT. (c), (d), (e) are examples of the different unit cells of metamaterial CIT devices with different structures of bright modes and dark modes. The geometrical parameters of (c) are $W = 10 \mu m $, $L = 80 \mu m$, $d = 4 \mu m$, $w = 4 \mu m$, $l_x = 26 \mu m$ $l_y = 28 \mu m$, $P_x = P_y = 100 \mu m$ and $g_1 = 4 \mu m$ with varying gap $g_2$ of SRR. The geometrical parameters of (d) are $L_x = 92\mu m$, $L_y = 16 \mu m$, $l = 40 \mu m$, $d = 8 \mu m$, $w = 4 \mu m$ $P_x = P_y = 100 \mu m$ and $l_1 = 30 \mu m$ with varying side length $l_2$ of U shape.  The geometrical parameters of (c) are $W = 10 \mu m$, $L = 92 \mu m$, $w = 5 \mu m$, $d = 5 \mu m$, $g = 50 \mu m$, $P_x = 100 \mu m$, $P_y = 120 \mu m$ and $R_1 = 20 \mu m$ with varying $R_2$.}
	\label{Fig1}
\end{figure}

It is worth emphasizing at the beginning that CIT may be easily confused with another phenomenon called electromagnetically induced transparency (EIT). The transmission spectrum of CIT looks similar to EIT and both phenomena come from bright mode and dark mode coupling. However, CIT is totally different from EIT. Normally, EIT comes from the coupling between bright mode and uniform dark mode (the dark mode has the same geometrical parameters). The most remarkable part of CIT is to propose two slightly different dark modes. Thus, in contrast to EIT, we should consider the interference between the dark modes in CIT. To sum up, the bright mode of EIT directly couples to the the dark modes. However, the bright mode of CIT couples to the the interference of the dark modes. Therefore, two-level coupled mode theory (CMT) is enough to describe EIT and we should employ three-level CMT to well describe CIT due to considering interference between the dark modes. Thus, the physics of our proposal is entirely the same as the physical origin of CIT in quantum system \cite{Lei2023}, but it is quite different from EIT. 
Note that, we produce the transmittance dip with the asymmetric dark modes. Strictly speaking, transmittance dip can not be defined it as transparency. However, the terminology CIT comes from the paper \cite{Lei2023} which produces the dip in the reflection spectrum. We produce an exactly similar dip in the transmission spectrum. Due to the equivalent physics, we just follow the terminology CIT in our transmission spectrum. 

Due to the equivalent analogue of quantum optics \cite{Huang2022}, the transmission spectrums of the coupling between metamaterial structures can effectively analogize many quantum effects, such as EIT \cite{Papasimakis2008, Chiam2009, Liu2010, Gu2012, Zhang2008, Yahiaoui2018,  Xiong2009, Yang2014}, bound states in the continuum (BIC) \cite{Cong2019, Huang2021, Tan2021, Fedotov2007, Koshelev2018, Abujetas2019, Koshelev2019, Kupriianov2019, Huang20231, Tan2020, Wang2023}, PT-symmetry \cite{Droulias2019, Lawrence2014} and employing quantum control \cite{Huang2019, Huang2020, Huang20212}. In this paper, we propose the analogue of CIT in coupling of metamaterial structures for the first time, as shown in Fig. \ref{Fig1}. In this paper, we demonstrate the three different types (Fig. (c), (d) and (e)) of coupling of bright modes (cut wire in (c), (e); rectangle in (d)) and dark modes (split-ring resonator in (c); U shape in (d); C shape in (e)), where bright mode and dark mode are analogue to cavity and collective ions respectively. The reason why we choose different structures is that we want to demonstrate our idea is not limit to special metamaterial structures and our theory is universality. 

\begin{figure}[htbp]
	\centering
	\includegraphics[width=0.45\textwidth]{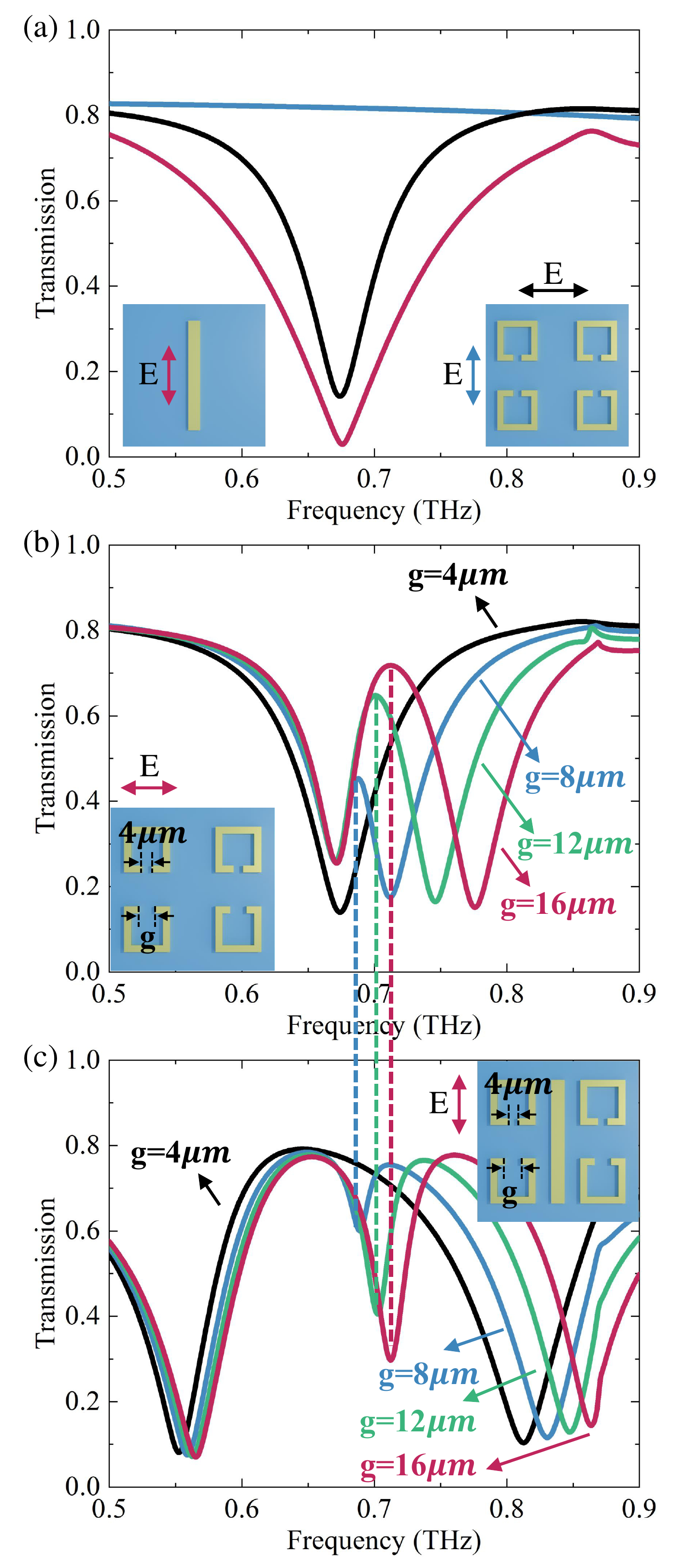}
	\caption{(a) The black line is shown as the bright mode of CW with $x$ polarization. The blue line and the black line are shown as the dark mode with $x$ polarization and $y$ polarization respectively.  (b) The dark modes with $y$ polarization when we slightly vary the geometrical parameter of one bright mode (gap $g_1$ and $g_2$ in this example.). (c) The corresponding CIT transmission spectrum with different gap $g$ of dark modes. }
	\label{Fig2}
\end{figure}

The physical meaning of bright mode is that the metamaterial structure can be excited by external wave, and dark mode can not excited by external wave. However, the dark mode can be excited by bright mode due to the coupling between bright mode and dark mode. 
In this paper, we do not try to design novel metamaterials' structures and we plan to employ the some typical structures which can well demonstrate our idea. Thus, the bright mode of metamaterial structure (cutting wire (CW) or rectangle) is coupling to the dark modes of metamaterial structures (split-ring resonators (SRRs) or U shapes or C shapes), which is effectively equivalent to the coupling of the cavity and collective ions. Although there are a large number of ions as the collective ions in cavity QED system and our dark modes of metamaterial structures are only configured as two structures, the main contribution of collective ions to CIT is to produce the quantum interference between the collective ions. However, we can slightly vary geometrical parameter of one dark mode structure to make it slightly different from another dark mode structure. Therefore, due to the symmetry broken, two dark mode structures also can generate interference of each other as quasi-BIC, which is exactly the same as the interference of collective ions. 

The motivations for introducing CIT into metamaterial are that \textit{i)} CIT is a new optical phenomenon due to the interference between ions within the cavity, which was first demonstrated in the quantum cavity-ions system \cite{Lei2023}. We find the equivalent interference between the two different dark modes to realize CIT in the classical metamaterial system. Our device is much easier implementation than quantum system. \textit{ii)} Besides, our paper provides a new mode of light-matter interaction in the metamaterial and brings a deeper understanding of metamaterial coupling. \textit{iii)} We conclude and summarize the bright mode with coupling interference of all different dark modes as the CIT effect and clearly understand physics beyond this whole kind of configuration metamaterial. Furthermore, ideal CIT metamaterial can provide a new very narrow-band peak within EIT peak, which could be very useful for THz sensor \cite{Tao2011, Shen2022}, high-sensitive optical switching and  steering of terahertz communication \cite{Cong2020}. 
Few pioneer papers have already discussed two different dark modes in the metamaterial EIT \cite{Zhang2021, Sarkar2019, Hu2022}. However, those studies do not realize that it is a brand new optical phenomenon and illustrate it as a transformation of EIT, such as the multi-band EIT \cite{Zhang2021, Sarkar2019} and EIT-like effect with considering BIC \cite{Hu2022}. Therefore, previous theories only employ coupled mode theory (CMT) to fit the transmission spectrum and lack of understanding physics behind it. 

In this paper, we firstly demonstrate how to establish CIT in metamaterial and we give the theoretical analysis by CMT, which is consistent with our proposal. Furthermore, we obtain the CIT phenomena with different bright modes and dark modes to claim that our theory is universality. Finally, we do the experiments with measuring transmission spectrums by THz Time-domain system (TDS). Due to the limitations of large lossy of metal-metamaterial and resolution of our THz TDS, we only can obtain low Q peak of CIT in our demos. However, we still can obtain very clear tendency of CIT in metamaterial to well verify the corrections of our theory, analysis and simulations. We do not expect that our demos can perform as a real functional devices, but our demos open the approach of CIT in metamaterial for the first time.

\section{Results}
\subsection{Collectively induced transparency in metamaterial}
In one specific example (Fig. \ref{Fig1} (c)), the bright mode (CW) can be considered as the cavity and the dark modes can be considered as the collective ions. Assume the polarization of exciting THz wave is $x$ direction and external THz wave only can excite the bight mode (CW) as shown in red line of Fig. \ref{Fig2} (a) and can not excite the dark modes (SRRs), as shown in blue line of Fig. \ref{Fig2} (a). The CW can excite the dark mode with $y$ polarization, thus, we should give the full-wave simulation for dark modes without bright mode with $y$ polarization, as shown in black line of Fig. \ref{Fig2} (a). 
The more interesting is that when slightly vary the geometric parameters of one dark mode (gap $g_2$ in this example), two different structures of dark modes could have the interference effect, which is the well-known symmetry-breaking BIC in metamaterials, as shown in the blue, green and red lines of Fig. \ref{Fig2} (b) and two interference of dark modes are corresponding to the quantum interference of collective ions. 

It is no surprising that if there is no symmetry-break in the dark mode (see the black line of Fig. \ref{Fig2} (b)), the energy of bright mode can flow to the dark mode due the coupling between the bright mode and dark modes, and after that there is no energy remaining in the bright mode, which is common metamaterial EIT, as shown in the black line of Fig. \ref{Fig2} (c).  
It is remarkable that when two metamaterial structures are slightly different, the metamaterial BIC starts to appear with corresponding to the blue, green and red lines of Fig. \ref{Fig2} (b). The typical transmission spectrum of quasi-BIC is the Fano shape and the overall dark modes become suppressed due to the interference of dark modes at the maximum point of Fano shape. Thus, the overall dark modes can not be excited by bright mode at the maximum frequency of Fano shape and it leads that the coupling between bright mode and dark modes becomes much smaller. 
Therefore, the energy of bright mode can not flow to dark modes due to the suppression of dark modes. Consequently, bright mode can excite by external THz wave again, which causes the CIT effect, as shown in the blue, green and red lines of Fig. \ref{Fig2} (c).  As we can see from the results, all the maximum frequencies of Fano shape in BIC in Fig. \ref{Fig2} (b) correspond to CIT frequencies in Fig. \ref{Fig2} (c) respectively and these results verify our analysis based on the coupling between the bright mode and the interference of dark modes. 

It is worth emphasizing that as we can see from the blue line of Fig. \ref{Fig2} (a), the SRR is dark for the y-polarization external field. When we slightly vary the gap of SRR, we believe that the resonant frequency of SRR is only slightly shitting, thus it should be still dark for the y-polarized external field. When the incident polarization is parallel to the SRR gap, the SRRs (dark modes) become bright, because the bright mode (CW) can provide an equivalent x-polarized field to excite the dark modes. Therefore, the y-polarized external THz wave only can excite the bright mode (CW), but it can not excite the dark modes (SRRs). The dark modes (SRRs) can be excited due to the coupling of bright mode and dark modes. Meanwhile, two slightly different SRRs start to interfere with each other and this interference also affects the transmission spectrum.

\subsection{Theoretical derivation}
Beyond the qualitative analysis of CIT metamaterial, we can employ the coupled mode theory (CMT) to demonstrate the CIT metamaterial in analytical solution. Based on the CMT, the coupling between one bright mode and two different dark modes can be written as,

\begin{widetext}
\begin{equation}
\left[
	\begin{matrix}
	\omega - \omega_b - i\gamma_b & g_1 & g_1 \\
	g_1 & \omega -  \omega_d - i\gamma_d & g_2  \\
	g_1 & g_2 & \omega - (\omega_d + \Delta_{\omega}) - i(\gamma_d + \Delta_{\gamma})%
	\end{matrix}%
	\right] \left[
	\begin{matrix}
	a \\
	b \\
	c %
	\end{matrix}%
	\right] =\left[
	 \begin{matrix}
	 \sqrt{\gamma_b} \\
	 0 \\
	 0%
	 \end{matrix}%
	 \right]E ,
\end{equation}%
\end{widetext}
where $|a|^2$, $|b|^2$ and $|c|^2$ are the energies in each metamaterial structure and $\omega$ is the frequency of input THz wave. $\omega_b$ is the resonant frequency of bright mode (CW) with $x$ polarization and $\gamma_b$ is the loss rate of bright mode (CW). $\omega_d$ and $\omega_d + \Delta_{\omega}$ are the resonant frequencies of dark modes (SRRs) with $y$ polarization and corresponding loss rates are $\gamma_d$ and $\gamma_d + \Delta_{\gamma}$. Due to the slight changes in geometric parameter of one dark mode, the resonant frequency and loss become slight different. $g_1$ is the coupling strength of bright mode and dark modes. $g_2$ is the coupling strength of bright modes.

Due to the best coupling from EIT, the normal resonant frequency of bright mode with $x$ polarization $\omega_b$ is the same as the resonant frequency of dark mode with $y$ polarization $\omega_d$, notation as $\omega_b \approx \omega_d = \omega_0$. To simplify our calculations, we assume the loss rate of bright mode and dark mode are nearly the same, notation as $\gamma_b \approx \gamma_d = \gamma$. Compared with the coupling strength of bright mode and dark modes $g_1$, coupling strength of dark modes $g_2$ is relatively much smaller. Thus, we can ignore the $g_2$ and set $g_1 = g$. The transmission spectrum can be calculated as $T \approx 1 - Im(\chi_{\text{eff}})$ \cite{Huang2021, Huang2019, Huang2024}, where $\chi_{\text{eff}} = \frac{\sqrt{\gamma} a}{E}$. By solving the CMT (Eq. 1), we can obtain the $\chi_{\text{eff}}$, which is

\begin{figure*}[htbp]
	\centering
	\includegraphics[width=0.8\textwidth]{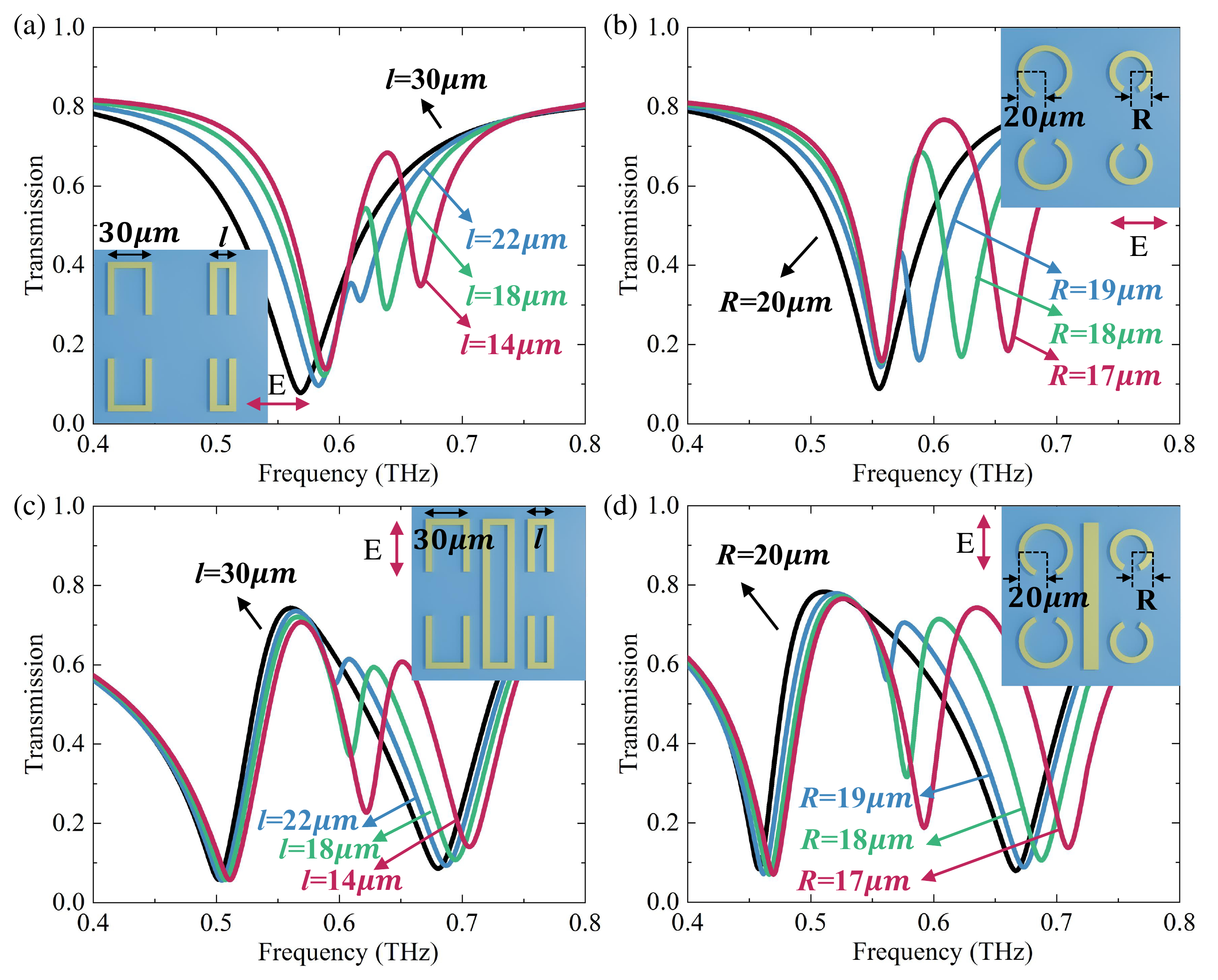}
	\caption{The metamaterial CIT with different bright mode and dark modes. (a) The transmission spectrums of dark mode (U shape) with $y$ polarization by varying different side lengths $l$ of one dark mode. (b) The transmission spectrums of dark mode (C shape) with $y$ polarization by varying different radius $R$ of one dark mode. (c) The coupling of bright mode (rectangle) and interference of dark modes (U shape). (d) The coupling of bright mode (CW) and interference of dark modes (C shape).} 
	\label{Fig3}
\end{figure*}

\begin{widetext}
\begin{equation}
\chi_{\text{eff}} \approx \dfrac{-i \gamma (i \gamma - \Omega) (i \gamma + i \Delta_{\gamma}-\Omega + \Delta_{\omega})}{g^2\Delta_{\gamma} - i g^2 \Delta_\omega + \Delta_{\gamma} \gamma^2 - i \Delta_{\omega} \gamma^2 + 2 g^2 \gamma - \Delta_{\gamma} \Omega^2 + i \Delta_{\omega} \Omega^2 + 2i g^2 \Omega - 3 \gamma \Omega^2 + 3i \gamma^2 \Omega + \gamma^3 - i \Omega^3 + 2i \Delta_{\gamma} \gamma \Omega + 2 \Delta_{\omega} \gamma \Omega};
\end{equation}
\end{widetext}
where $\Omega = \omega - \omega_0$. Note that $g$, $\gamma$, $\Delta_{\omega}$, $\Delta_{\gamma}$ are relatively smaller comparing with $\Omega$. Therefore, we can ignore the high-order minimum and then we can obtain the approximation of $\chi_{\text{eff}}$, which is 
\begin{equation}
\chi_{\text{eff}} \approx \dfrac{i \gamma}{(3\gamma + \Delta_{\gamma}) + i(\Omega - \Delta_{\omega})}
\end{equation}
Consequently, the transmission spectrum of CIT metamaterial is,
\begin{equation}
T \approx 1 - \dfrac{3\gamma^2 + \gamma \Delta_{\gamma}}{(\omega - \omega_0 - \Delta_{\omega})^2 + (3 \gamma + \Delta_{\gamma})^2}.
\end{equation}
We find the transmission spectrum of CIT metamaterial as a function of a Lorentzian Profile: 
\begin{equation}
T = 1 - \dfrac{A}{(\omega - \omega_{CIT})^2 + (\Delta_{CIT}/2)^2},
\end{equation}
where $\omega_c$ is the center frequency of CIT and $\Delta_{CIT}$ is the full width at half maxima (FWHM) of CIT shape, where
\begin{equation}
\omega_{CIT} \approxeq \omega_0 + \Delta_{\omega}; 
\end{equation}
\begin{equation}
 \Delta_{CIT} \approxeq 6 \gamma + 2\Delta_{\gamma}. 
\end{equation}
When we have more unsymmetric geometrical parameter of dark modes, $\Delta_{\omega}$ and $\Delta_{\gamma}$ are increasing. Therefore, the center frequency of CIT shifts to the higher frequency and FWHM of CIT also becomes larger, which is consistent to our Full-wave simulations in Fig. \ref{Fig2} (c). 
When two dark modes are identical, there is no opposed modes for two dark modes and it leads no interference between the dark modes, which reduces to typical metamaterial EIT. 


\subsection{Universality}
As we described in the theory of CIT metamaterial, there is no limit to the metamaterial structures for bright mode or dark modes. Therefore, our CIT metamaterial theory is universal for every kind of metamaterial structures. Therefore, we can switch the bright mode and dark modes to other metamaterial structures and we can obtain the same CIT effect in metamaterial based on our theory. Therefore, we can change other types of coupling between bright mode and the interference of dark modes to perform metamaterial CIT, as shown in Fig. \ref{Fig3}. We take rectangle and CW as the bright mode for Fig. \ref{Fig3} (a), (c) and Fig. \ref{Fig3} (b), (d), and we employ U shapes and C shapes as the dark modes for Fig. \ref{Fig3} (a), (c) and Fig. \ref{Fig3} (b), (d), respectively. We demonstrate the dark modes with $x$ polarization to represent the exciting dark modes from bight mode due to coupling, as shown in Fig. \ref{Fig3} (a) and (b). When we slightly vary the geometrical parameter of dark modes (side length $l$ of U shape and radius $R$ of C shape respectively), the metamaterial quasi-BIC starts to appear due to the symmetry-breaking, as shown in Fig. \ref{Fig3} (a) and (b). When we introduce the bright mode, the coupling between the bright mode and the interference of dark modes clearly causes the transmission spectrums from EIT to CIT, as shown in Fig. \ref{Fig3} (c) and (d). In addition, the maximum frequencies of Fano shape of dark modes are very close to frequencies of metamaterial CIT. Furthermore, we reduce the side length $l$ and radius $R$ which leads the higher resonant frequencies and larger $\Delta_{\omega}$. Hence, the frequencies shiftings of CIT are from lower to higher frequency, which is consistent with our analytical solution.

In addition, as we mentioned the physical origin of CIT metamaterial, CIT comes from the coupling of bright mode and the interference between the dark modes, and the bright mode is not directly coupling to the dark mode, but coupling to the interference between the dark modes. Overall, the CIT requires two conditions. Firstly, we only require two slightly different dark modes can be exited by bright mode and the interference of dark modes happens naturally. Furthermore, to make sure the coupling of bright mode and the interference of dark modes can happen. Therefore, we do not require a near-field coupling between dark modes, such as those systems \cite{Gupta2020, Singh2009, Tian2010}. In other words, if we let those designs become the dark modes and carefully design the bright mode. We also can provide CIT emerging from those systems \cite{Gupta2020, Singh2009, Tian2010}.

\begin{figure*}[htbp]
	\centering
	\includegraphics[width=0.4\textwidth]{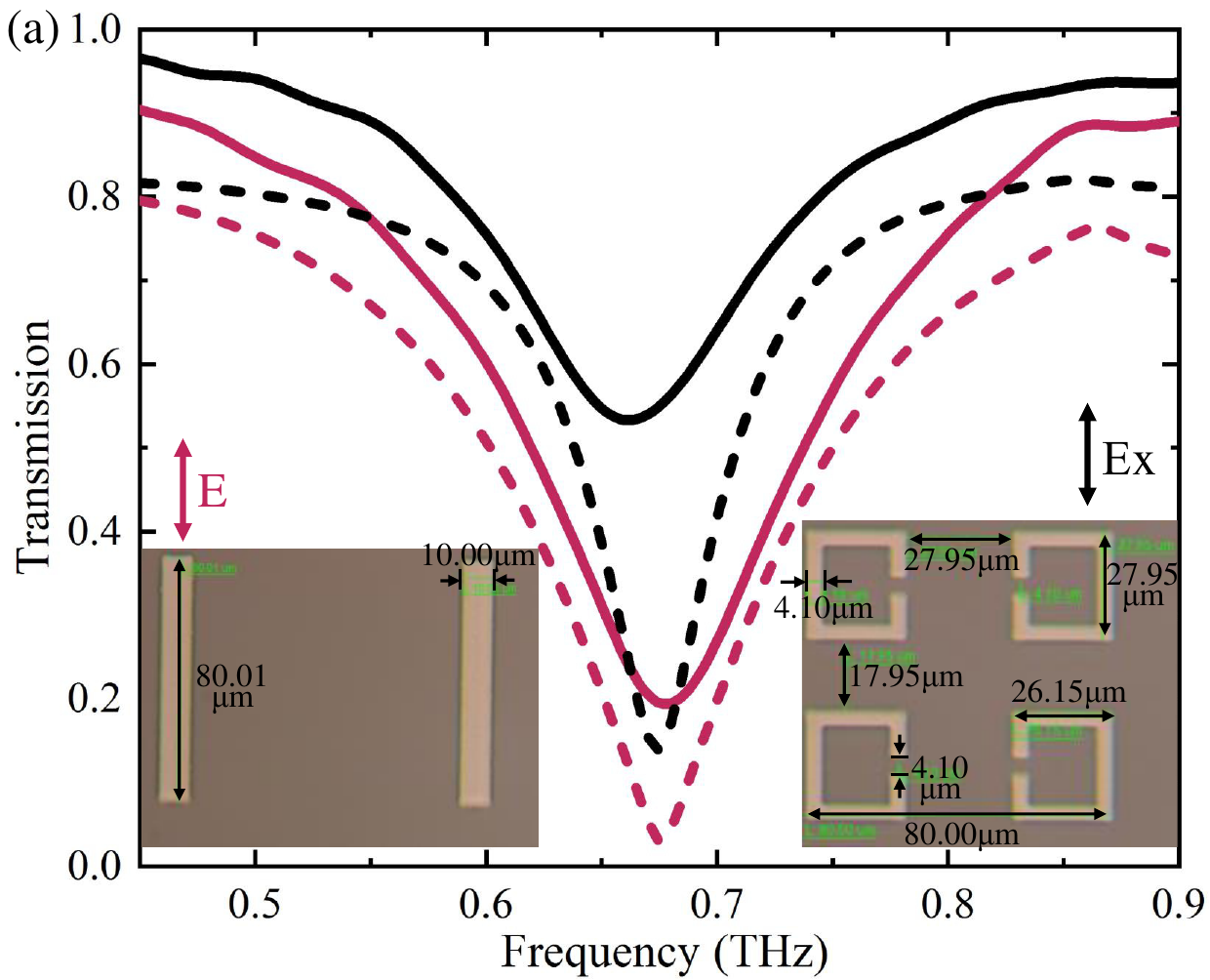}
	\includegraphics[width=0.4\textwidth]{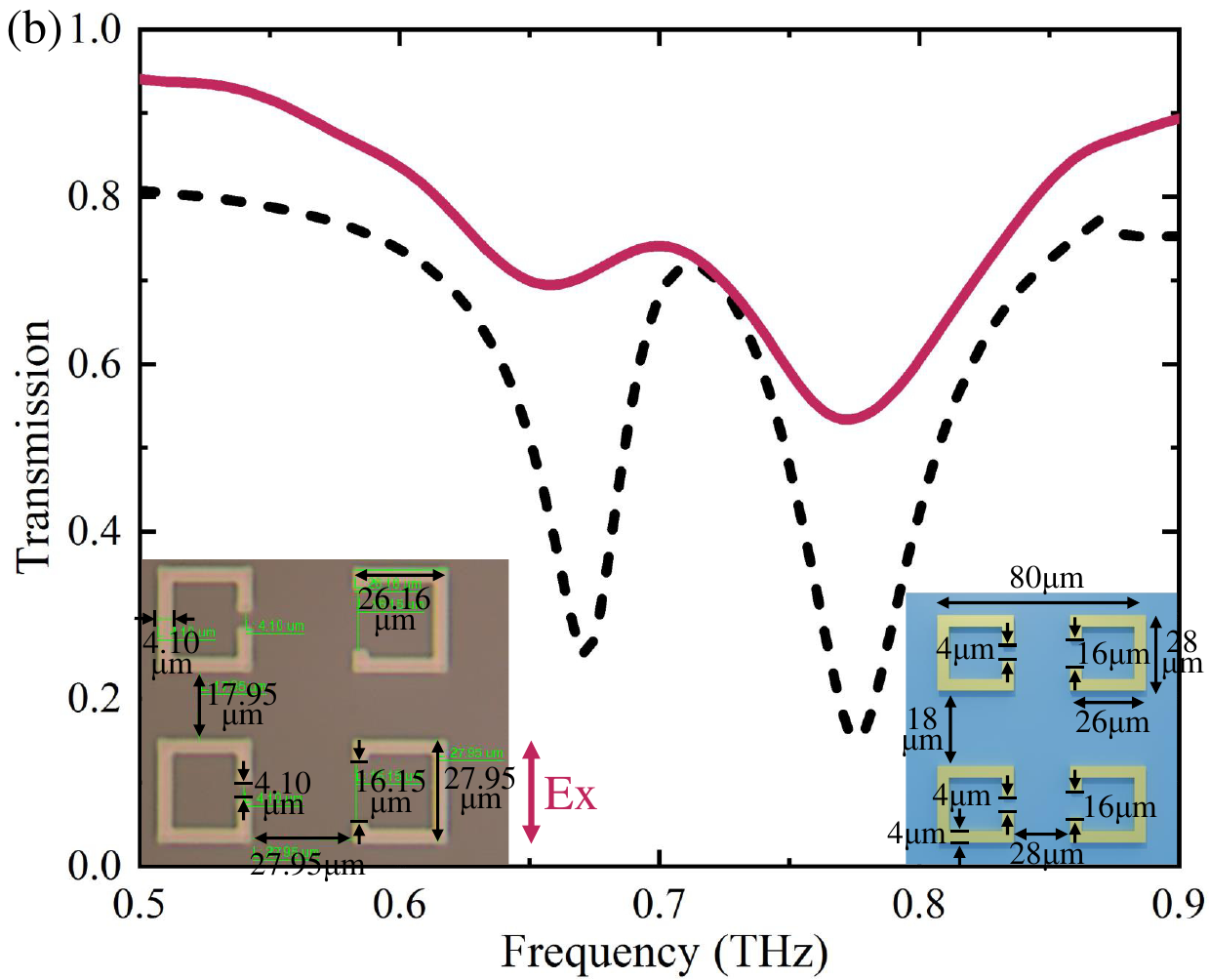}
	\includegraphics[width=0.4\textwidth]{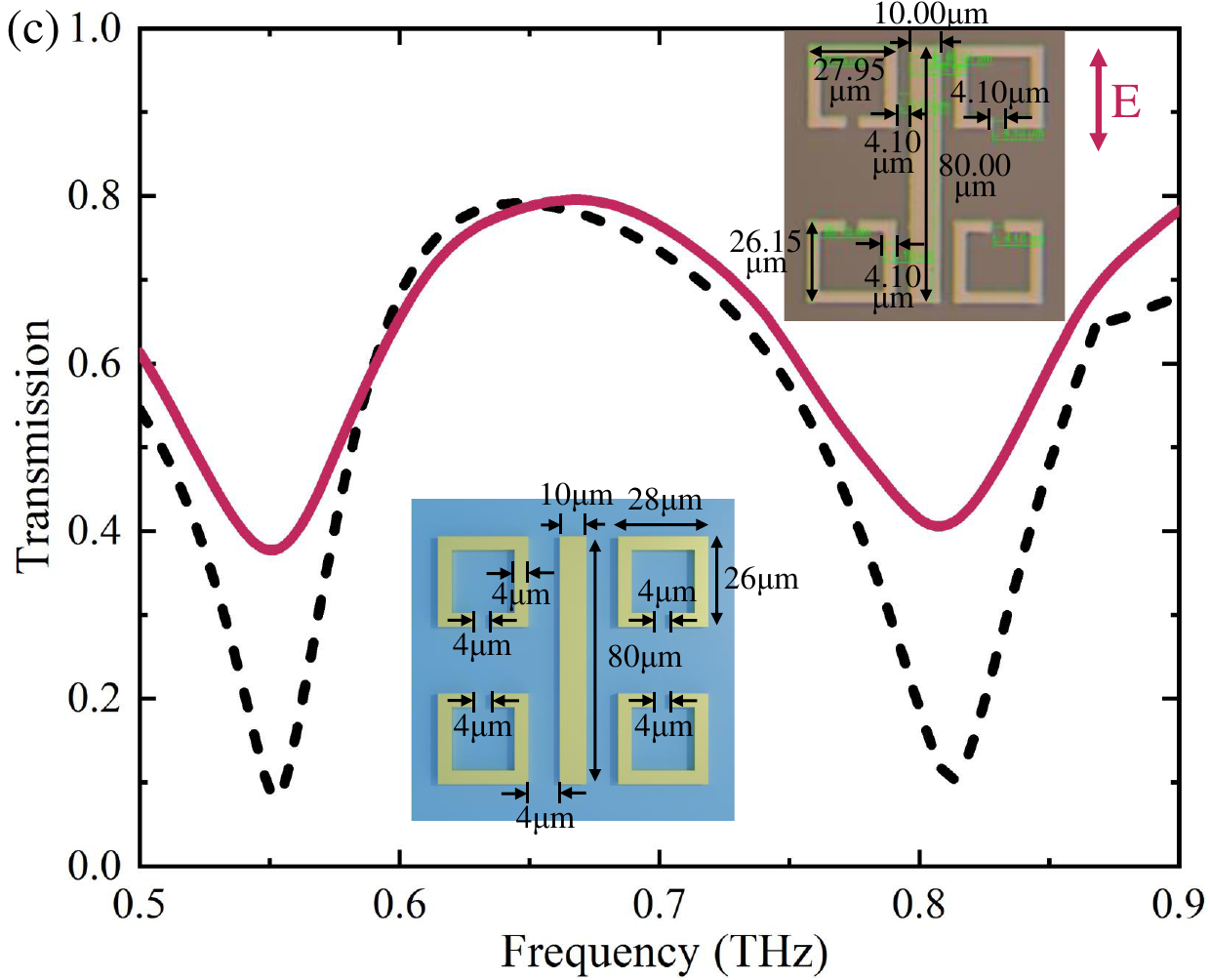}
	\includegraphics[width=0.4\textwidth]{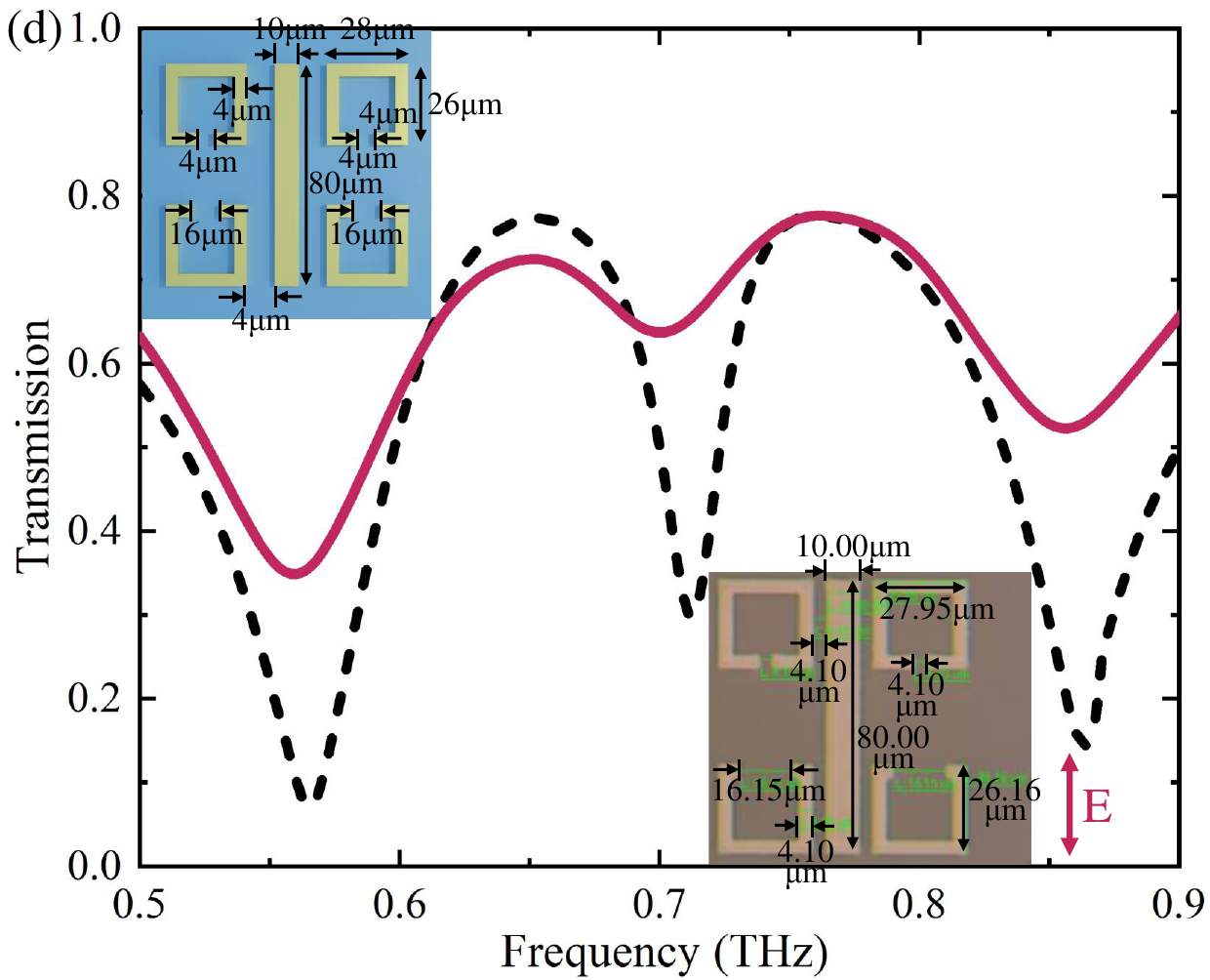}
	\includegraphics[width=0.4\textwidth]{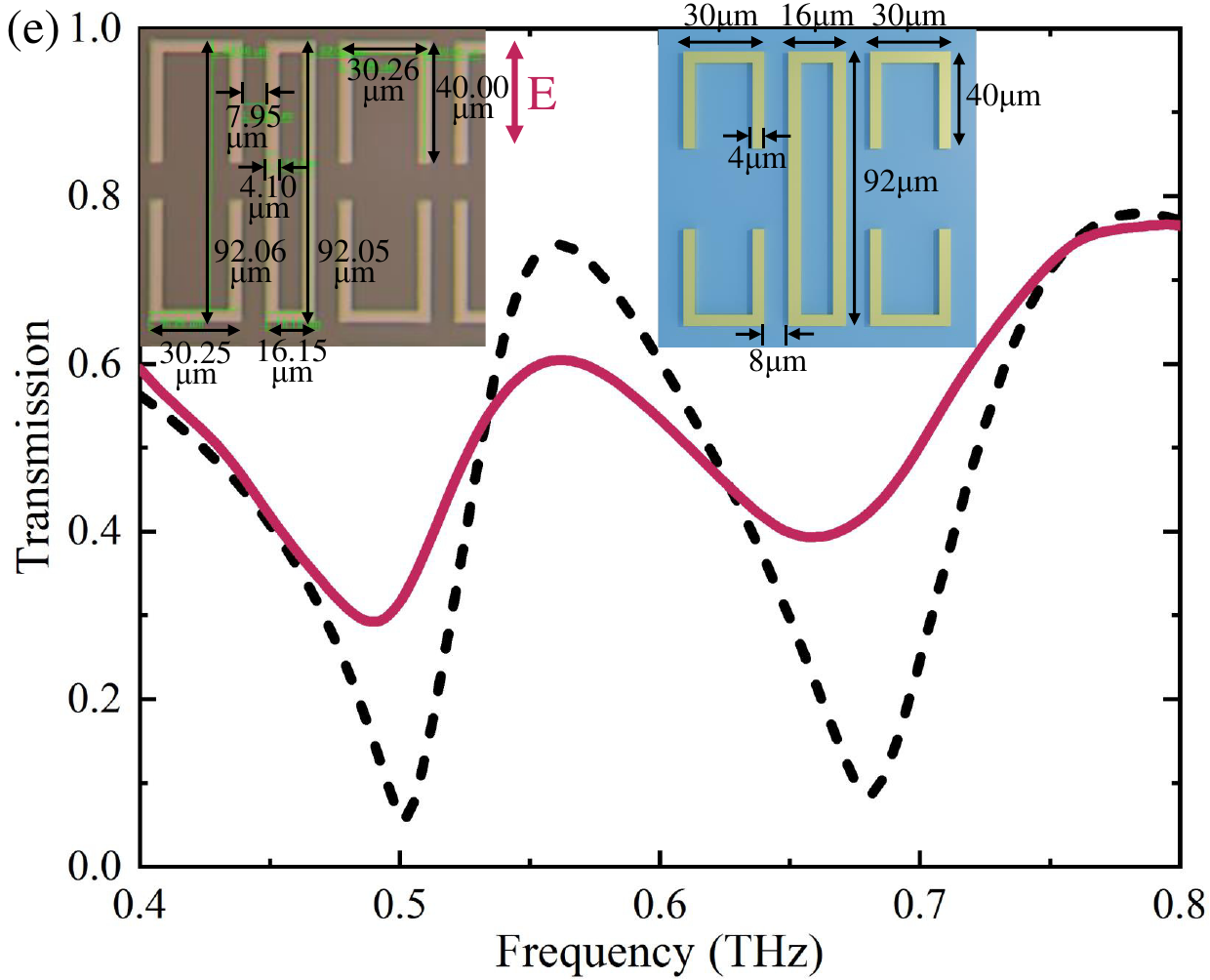}
	\includegraphics[width=0.4\textwidth]{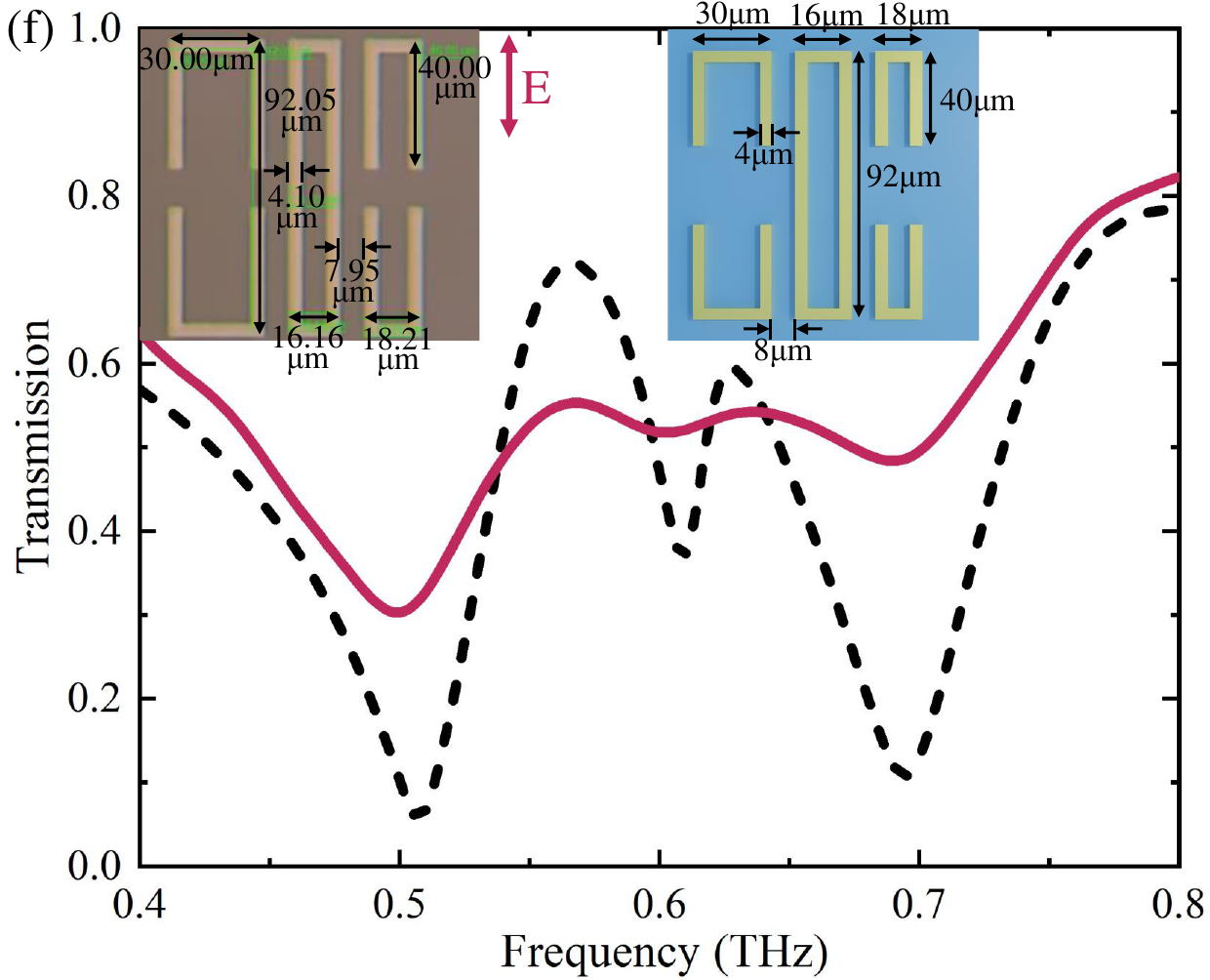}
	\includegraphics[width=0.4\textwidth]{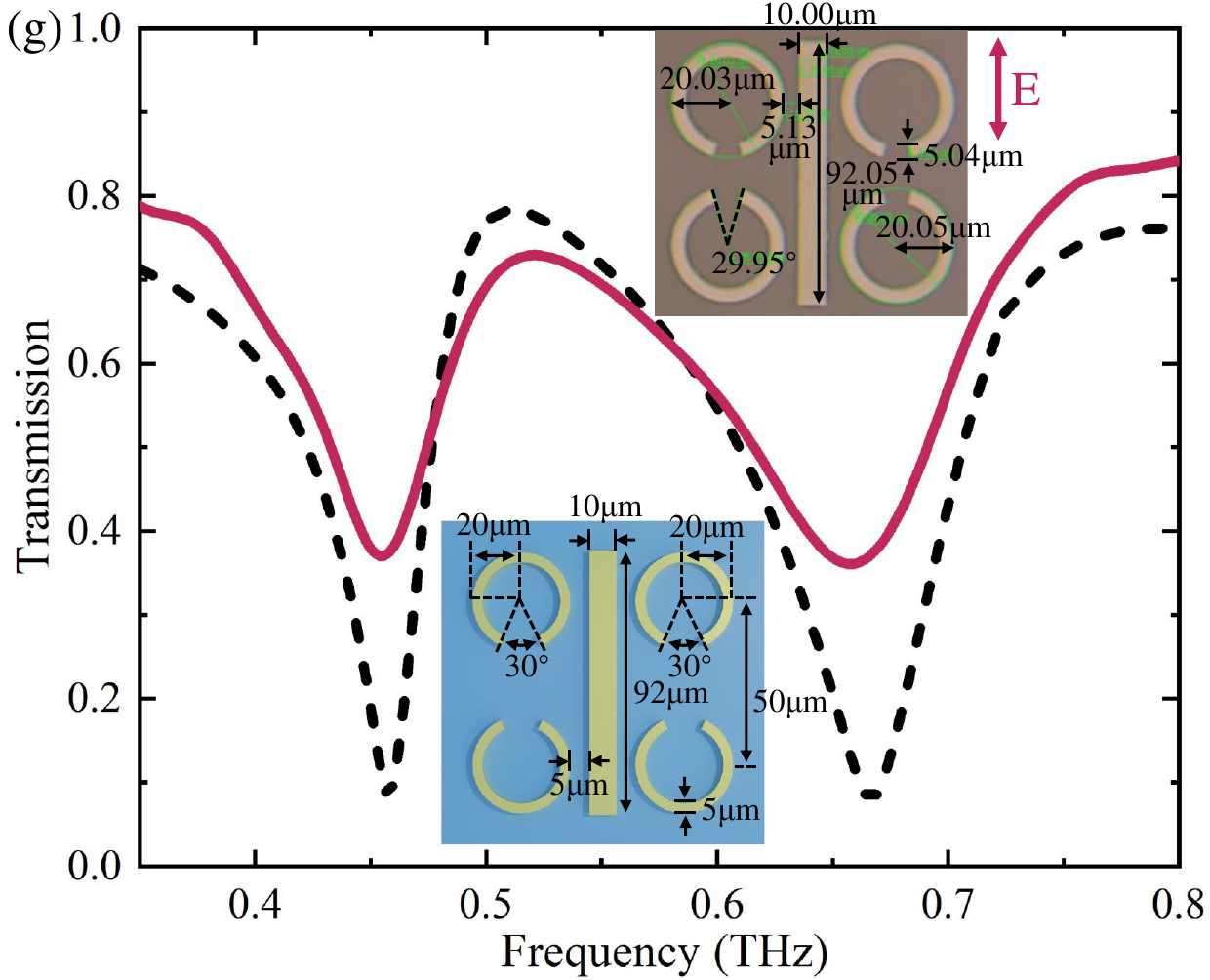}
	\includegraphics[width=0.4\textwidth]{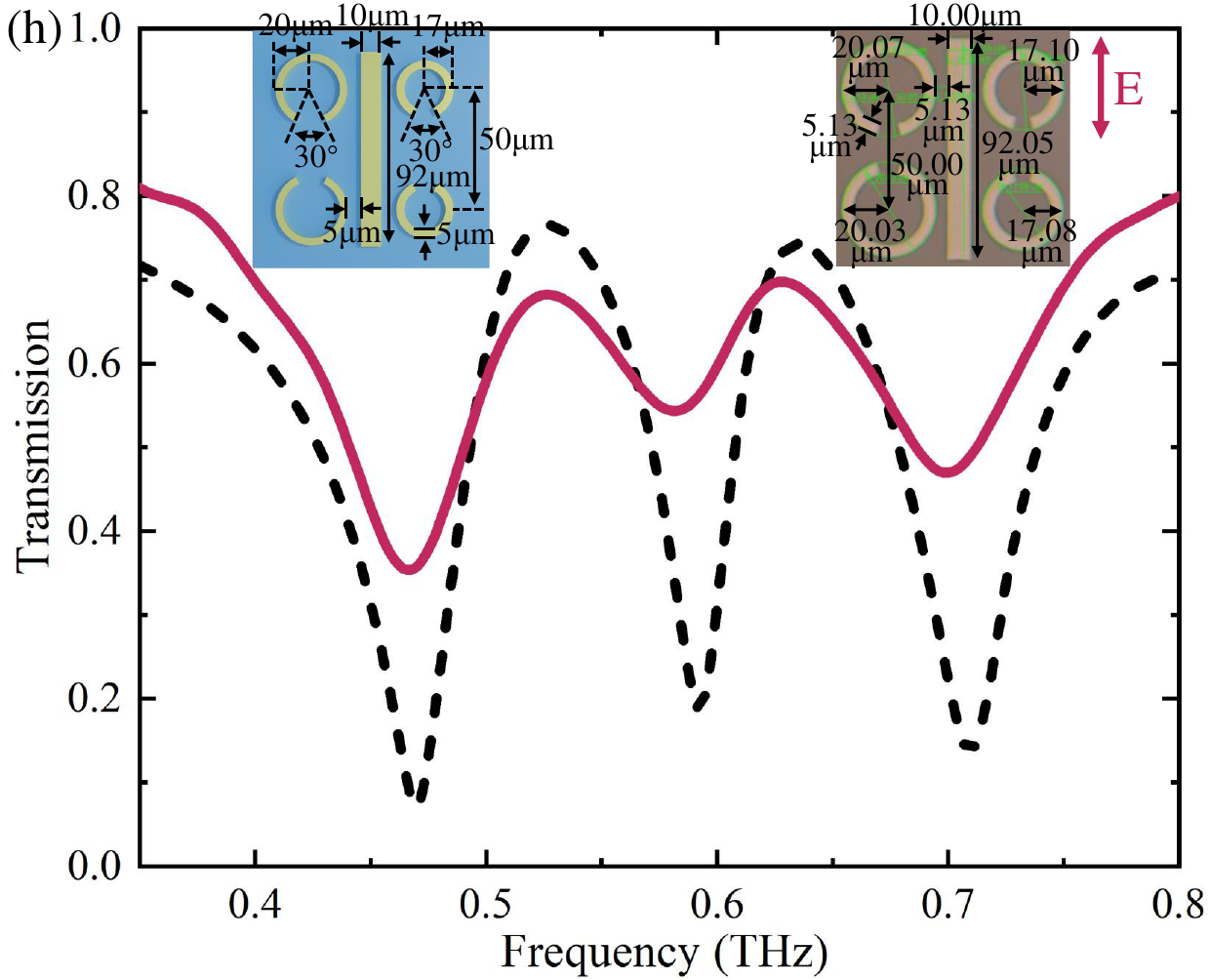}
	\caption{The experimental results of CIT metamaterial and the solid lines and dashed lines are the experimental results and simulations respectively. (a), (b), (c) and (d) are with respect to Fig. \ref{Fig1} (c) metamaterial.  (e) and (f) are with respect to Fig. \ref{Fig1} (d) metamaterial. (g) and (h) are with respect to Fig. \ref{Fig1} (e) metamaterial.}  
	\label{Fig4}
\end{figure*}

\section{Experiments}
In order to verify the validity of our theory and simulation, we demonstrate the experimental results to illustrate the corrections of our simulations by employing THz time-domain system, as shown in Fig.\ref{Fig4}.  Fig. \ref{Fig4} (a) demonstrates the bright modes (red lines) with $x$ polarization and dark mode (black lines) with $y$ polarization. Fig. \ref{Fig4} (b) demonstrates the interference of dark modes with symmetry-breaking of gaps (4 $\mu m$ and 16 $\mu m$) of SRRs, which represents the quasi-BIC shape. Fig. \ref{Fig4} (c) and (d) represent the EIT and CIT with symmetric dark mode and un-symmetric dark modes. Furthermore, we demonstrate two different configurations of metamaterial to prove universality of our theory. Fig. \ref{Fig4} (e) and (f) are with respect to Fig. \ref{Fig1} (d) metamaterial. Fig. \ref{Fig4} (g) and (h) are with respect to Fig. \ref{Fig1} (e) metamaterial. (e) and (f) are the EIT with symmetric dark mode, corresponding to different metamaterials and (g) and (h) are the CIT with unsymmetric dark modes, corresponding to different metamaterials. 

As we can see from our experimental results, all the experimental results are well consistent with our simulations in the frequencies. Therefore, we can easily conclude that our simulations are correct with the verification of experiments. 
The deviation of amplitudes of transmission spectrums between the simulations and the experiments comes from two reasons. Firstly, the larger intrinsic loss of metal appears in the experiments especially for complex structures and also the loss of coupling between structures is not considerable in the simulations. Secondly, the thickness of substrate (Si) is only 1000 $\mu m$ and the more information of peak resonance will submerge within the echo when we do the TDS experiments.
The principle of time-domain system is that we measure the THz signal in time-domain and then we employ the fourier transform to calculate the transmission spectrum. However, when the THz wave passes through the substrate (Si), some THz wave will be reflected by this interface. The reflected THz wave will cover the original signal of the THz wave, which becomes the noise. In the experiments, we should discard the signal before the reflected THz wave. Ideally, we should measure the all signal in time domain as long as possible and then employ the fourier transform. However, due to this limitation, we lose the information that signals in time domain after the reflected THz wave. We are forced to lose the high-Q dips’ information and this is the reason we could obtain better experimental results in the lower Q dip (Fig. \ref{Fig4} (a)) compared with the higher Q dip (Fig. \ref{Fig4} (d), (f), (h)). Due to those reasons, we can not obtain the excellent agreements between experimental and simulation results \cite{Yin2013, Yahiaoui2018}. However, those deviations are very common phenomena in the metal-based metamaterials in THz TDS experiments.

In summary, there has deviations on the amplitudes and accuracy on the frequencies. However, we still can obtain very clear tendency of transmission spectrums of CIT metamaterial in the experiments. These transmission spectrums of experiments can well demonstrate the corrections of our theoretical analysis, analytical calculations and simulations. Note that due to the intrinsic defect of meta-based metamaterial, we can not obtain very high Q of CIT peak. 
However, we only demonstrate the corrections of our idea and we illustrate the CIT phenomena in metamaterial for the first time in this paper, but we do not try to fabricate the real functional devices. However, if we design the all-dielectric-based metamaterial, we can obtain the much higher Q CIT peak for real functional device due to the less lossy of all-dielectric-based metamaterial. 

\begin{figure}[htbp]
	\centering
	\includegraphics[width=0.45\textwidth]{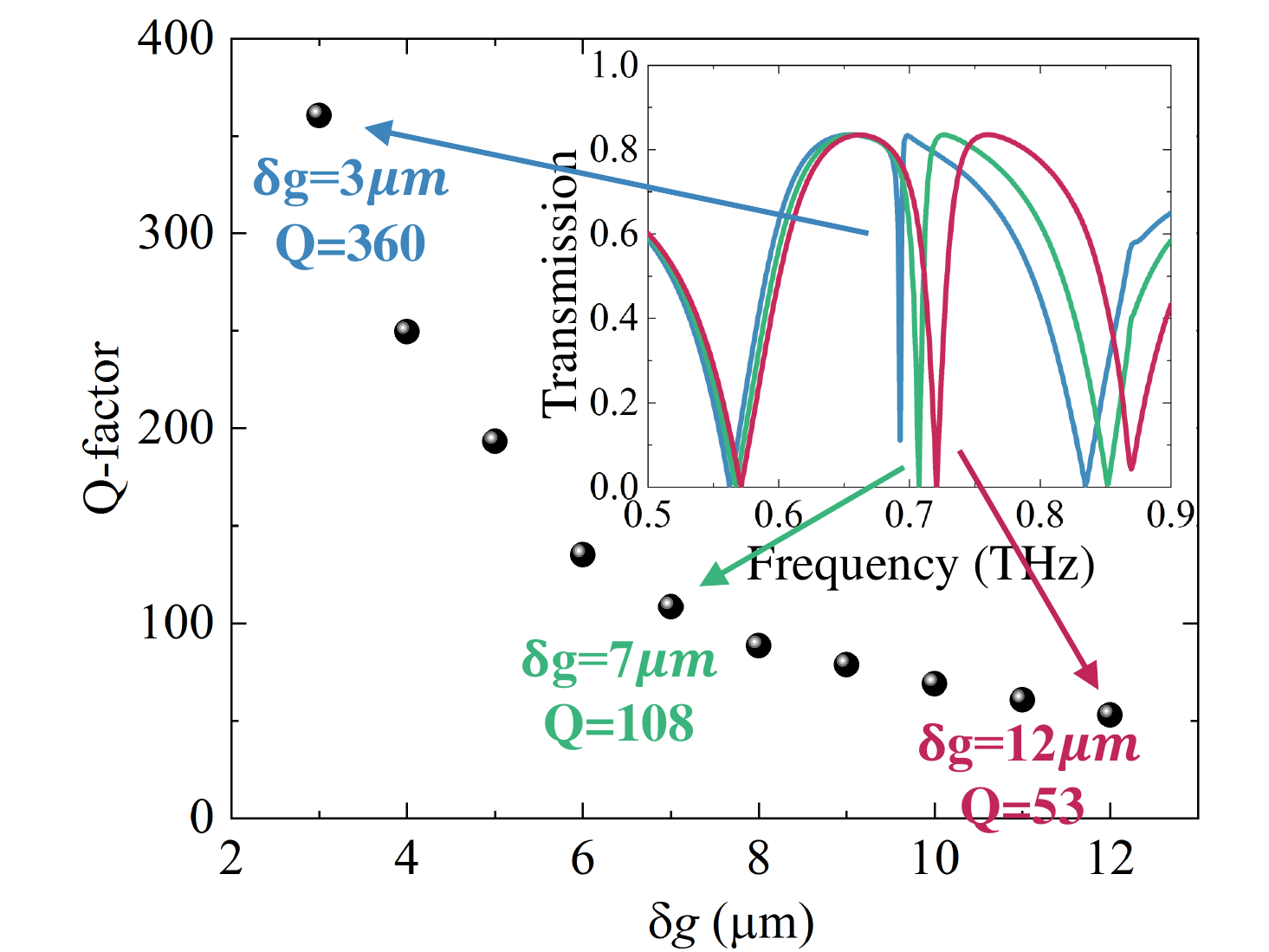}
	\caption{The simulations of ideal CIT metamaterial with changing aluminum to perfect electric conductor (PEC) based on metamaterial device as shown in Fig. \ref{Fig1} (c). The figure demonstrates the Q values of CIT with varying difference gap $\delta_g$ of dark modes, which is $\delta_g = g_2 - g_1$ and smaller figure demonstrates three typical transmission spectrums with $\delta_g = 3 \mu m, 7 \mu m, 12 \mu m$ respectively. }  
	\label{Fig5}
\end{figure}

As we mentioned that ideal CIT can produce very high-Q peak due to the limitations of our THz TDS system and metal-based metamaterial. However, we still can prove the correction of our idea by replacing metal to perfect electric conductor (PEC) in the CST simulations, as shown in Fig. \ref{Fig5}. In this example, we employ the CIT structures as shown in Fig. \ref{Fig1} (c) and we define the difference gap $\delta_g$ of dark modes, which is $\delta_g = g_2 - g_1$. From the results of simulations, we can easily obtain the very high-Q peak of CIT with ideal cases. 
Furthermore, Due to the dark modes are closer similarity (smaller $\delta_g$), the $\Delta_\gamma$ and $\Delta_\omega$ in Eq. 1 becomes smaller. Thus, it leads that frequency of CIT is closer to bight mode and smaller FWHM of CIT which causes the higher Q value, as shown in Eq. 6 and Eq. 7. From Fig. \ref{Fig5}, we can obtain higher Q value of CIT along with $\delta_g$ decreasing and frequencies of CIT and bright mode are getting closer. Our simulations are well consistent with prediction of our theory.

\section{Discussion}
It is worth mentioning that another way to present one more peak within EIT \cite{Liu2018, Yin2013} which is similar to our EIT. However, It is quite different from our CIT. They produce one more peak within EIT by employing symmetry-breaking of coupling between the bright mode and dark mode. Our CIT comes from the symmetry-breaking of dark modes to cause the interference of dark modes and then due to bright mode coupling with interference of dark modes, the transmission spectrum emerges the CIT phenomenon. 
The previous paper \cite{Yahiaoui2018} proposed the EIT from bright-bright modes coupling, which is similar to Fig. \ref{Fig2} (b). However, our CIT is demonstrated in Fig. \ref{Fig2} (c). Note that Fig. \ref{Fig2} (b) and (c) have different polarization. Therefore, in our paper, two dark modes can not be excited by external field without bright mode (CW). Thus, the physics behind CIT and EIT from bright-bright modes coupling are entirely different.

\begin{figure}[htbp]
	\centering
	\includegraphics[width=0.5\textwidth]{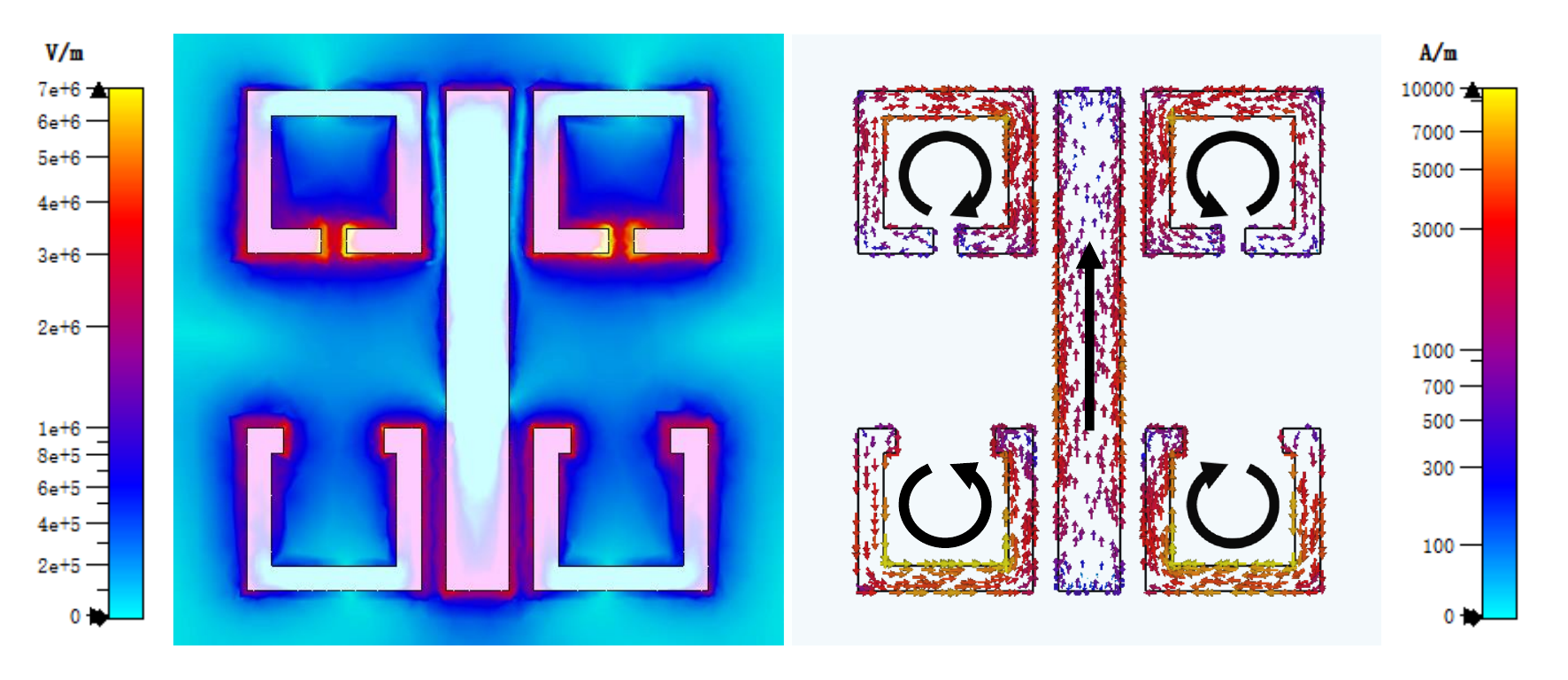}
	\caption{The electric field (left) and current density field (right) at CIT frequency,  with corresponding to our example as shown in the red line in Fig. 2 (c) at around 0.72 THz.}  
	\label{Fig6}
\end{figure}

For better understanding, We plot the electric field and current density field at CIT frequency, as shown in fig. \ref{Fig6} with corresponding to around 0.72 THz for the red line in Fig. \ref{Fig2} (c). As we can obtain from the results, the dark modes present two opposite modes which provide destructive interference between the dark modes at CIT frequency. Therefore, the bright mode can not couple to dark modes at this frequency due to the suppression of dark modes, which remains some current density and energies. Thus, the bright mode can be re-excited by external THz wave and it causes the CIT dip in the transmission spectrum.
The CIT dip comes from the interference of dark modes due to asymmetric dark modes. The interference of dark modes can provide the Fano shape in the transmission spectrum, as shown in Fig. \ref{Fig2} (b). Fig. \ref{Fig2} (b) demonstrates the transmission spectrum of dark modes with x-polarized THz wave without bright mode and this configuration is equivalent to the excitation of dark modes by bright mode’s coupling with y-polarized THz wave. This Fano shape of the transmission spectrum is the BIC shape due to the asymmetric dark modes. The physics behind BICs comes from the interference of different structures' radiated waves, providing the sub-radiated mode (destructive interference) and super-radiated mode (constructive interference) which causes the Fano shape. The sub-radiated mode suppresses the excitation of dark modes and it leads that the bright mode can not coupled to dark modes which causes the CIT dip in the transmission spectrum. Therefore, our CIT comes from the coupling between the bright mode and interference between the dark modes.

For our simulation setup, we performed numerical simulations using the commercially available electromagnetic full-wave simulation software, CST Studio Suite 2018. We modeled aluminum (Al) as a lossy metallic material with a thickness of 50 $\mu m$. The conductivity of aluminum was set to $3.56 \times 10^7S/m$, while the dielectric constant of the silicon substrate was 11.9. In these simulations,to capture the periodic nature of the structure, we applied unit cell boundary conditions in the x- and y-directions, and implemented an open boundary condition along the z-direction in free space. The incident THz waves were normal, and the polarization directions for different cases are detailed in the figure's caption. Subsequently, we utilized a tetrahedral mesh for the frequency domain solver to obtain the simulated THz transmission response.

For our experimental setup, we employ the conventional photolithography technology to fabricate our metamaterial samples with 200 $nm$ thickness of aluminium as the metal and 1000 $\mu m$ high resistance silicon as the substrate. Due to large geometrical parameters of structures (in units of $\mu m$), there is nothing trick on the fabrication which is a mature and reliable process. We employ the THz Time domain system (TDS) to measure the transmission spectrums of our simples, which is  all-fiber terahertz Time domain system (model TPF15 K) produced by Terahertz Photonics Technology. The THz waves are generated by commercial femtosecond laser with a central wavelength of 780 nm hitting on GaAs photoelectric crystal. After THz waves generation, we implement the THz lens to obtain 8-F confocal configuration and all the simples are placed on the focus point of THz lens system. Subsequently, the THz waves go through the simples and then hit on the ZnTe crystal to convert THz waves back to optical pulses. By combining optical pulses with passing through the simples and original optical pulses from femtosecond laser with delay line, the whole optical pulses passes through a quarter-wave plate and Wollaston prism. A photodiode converts the optical signal into an electrical signal, and then amplifies the signal through a lock-in amplifier. By applying Fourier transform, it is very easy to obtain the transmission spectrums and corresponding phase spectrums. The final transmission spectrums are given by  $t(\omega) = |E_S (\omega)/E_R (\omega)|$, where $E_S (\omega)$ and $E_R (\omega)$ are the Fourier transformed spectra of the transmitted sample signal and substrate pulses respectively.

\section{Conclusion}
In this paper, we propose the transmission spectrums of novel optical phenomenon, named collectively induced transparency (CIT) in the Terahertz metal metamaterial for the first time. Our CIT metamaterial based on the coupling between the bright mode and interference of dark modes by producing un-symmetrical dark modes. We give the analytical analysis and derive the analytical solutions of CIT based on coupled mode theory. Based on our theory, we give the simulations and experiments to demonstrate the corrections of CIT metamaterial. Furthermore, our theory is universal to all kinds of CIT metamaterial.

\section*{Acknowledgements}
This work acknowledges funding from National Natural Science Foundation of China (grant no: 12264010; 61965005; 62205077; 62365004; 62235013). W.H. acknowledges funding from Guangxi oversea 100 talent project; W.Z. acknowledges funding from Guangxi distinguished expert project.

\end{document}